\documentclass[mathleft
]{an}
\usepackage{graphicx}
\usepackage{times}
\graphicspath{{./fig/}{./png/}}

\usepackage{amssymb}
\usepackage{amsmath}

\newcommand{\ovec}[1]{{\mbox{\boldmath $#1$}}}
\newcommand\Rm{\mbox{\textit{$R_m$}}}
\newcommand{\ba}{\ovec{a}}

\newcommand{\bb}{\ovec{b}}
\newcommand{\bB}{\ovec{B}}

\newcommand{\be}{\ovec{e}}

\newcommand{\bscE}{\ovec{\boldsymbol{\cal{E}}}}
\newcommand{\bF}{\ovec{F}}
\newcommand{\bff}{\ovec{f}}
\newcommand{\bg}{\ovec{g}}

\newcommand{\bh}{\ovec{h}}
\newcommand{\bH}{\ovec{H}}
\newcommand{\bj}{\ovec{j}}

\newcommand{\bk}{\ovec{k}}

\newcommand{\bu}{\ovec{u}}
\newcommand{\bU}{\ovec{U}}
\newcommand{\bW}{\ovec{W}}
\newcommand{\bx}{\ovec{x}}

\newcommand{\bzo}{{\bf 0}}

\newcommand{\bmB}{\overline{\ovec{B}}}

\newcommand{\bmH}{\overline{\ovec{H}}}

\newcommand{\bmU}{\overline{\ovec{U}}}

\newcommand{\mB}{\overline{B}}

\newcommand{\mU}{\overline{U}}
\newcommand{\bxi}{\ovec{\xi}}
\newcommand{\bgamma}{\ovec{\gamma}}

\newcommand{\bpsi}{\ovec{\psi}}
\newcommand{\bomega}{\ovec{\omega}}
\newcommand{\bOmega}{\ovec{\Omega}}
\newcommand{\bnab}{\ovec{\nabla}}
\newcommand{\hu}{\hat{u}}
\newcommand{\hb}{\hat{b}}
\newcommand{\EQ}{\begin{equation}}
\newcommand{\EN}{\end{equation}}
\newcommand{\EQA}{\begin{eqnarray}}
\newcommand{\ENA}{\end{eqnarray}}
\newcommand{\p}{\partial}
\newcommand{\x}{\times}
\newcommand{\lan}{\langle}
\newcommand{\ran}{\rangle}
\def\bzo {{\bf 0}}
\def\mB {\overline{B}}
\def\mU {\overline{U}}
\def\dd {\mbox{d}}
\def\hb {\hat{b}}

\def\hu {\hat{u}}

\def\iu {\mbox{i}}

\def\cE {\cal{E}}

\begin{document}

\Pagespan{1}{}
\Yearpublication{2009}%
\Yearsubmission{2009}%
\Month{09}%
\Volume{???}%
\Issue{??}%

\title{Mean electromotive force proportional to mean flow in mhd turbulence}

\author{K.-H. R\"adler\inst{1}\fnmsep\thanks{Corresponding author:
  \email{khraedler@arcor.de}\newline}
\and  A. Brandenburg\inst{2,3}}
\titlerunning{Electromotive force proportional to mean flow}
\authorrunning{K.-H. R\"adler \& A. Brandenburg}
\institute{
Astrophysikalisches Institut Potsdam, An der Sternwarte 16,
D-14482 Potsdam, Germany
\and
NORDITA, AlbaNova University Center, Roslagstullsbacken 23, SE-10691
Stockholm, Sweden
\and
Department of Astronomy, AlbaNova University Center,
Stockholm University, SE-10691 Stockholm, Sweden
}

\received{... 2009}
\accepted{... 2009}
\publonline{later}


\abstract
{In mean-field magnetohydrodynamics the mean electromotive force due to velocity and magnetic field
fluctuations plays a crucial role.
In general it consists of two parts, one independent of and another one proportional to the mean magnetic field.
The first part may be nonzero only in the presence of mhd turbulence, maintained, e.g., by small-scale dynamo action.
It corresponds to a battery, which lets a mean magnetic field grow from zero to a finite value.
The second part, which covers, e.g., the $\alpha$ effect, is important for large-scale dynamos.
Only a few examples of the aforementioned first part of mean electromotive force have been discussed so far.
It is shown that a mean electromotive force proportional to the mean fluid velocity, but independent of the mean
magnetic field, may occur in an originally homogeneous isotropic mhd turbulence if there are nonzero correlations
of velocity and electric current fluctuations or, what is equivalent,
of vorticity and magnetic field fluctuations.
This goes beyond the Yoshizawa effect, which consists in the occurrence of mean electromotive forces proportional
to the mean vorticity or to the angular velocity defining the Coriolis force in a rotating frame and depends
on the cross-helicity defined by the velocity and magnetic field fluctuations.
Contributions to the mean electromotive force due to inhomogeneity of the turbulence are also considered.
Possible consequences of the above and related findings for the generation of magnetic fields
in cosmic bodies are discussed.}

\maketitle

\section{Introduction}

Mean--field magnetohydrodynamics has proved to be a useful tool for studying the behavior
of mean magnetic fields in turbulently moving electrically conducting fluids
(see, e.g., Moffatt 1979, Krause \& R\"adler 1980, Brandenburg \& Subramanian 2005).
Within this framework both the magnetic field $\bB$ and the fluid velocity $\bU$
are split into mean parts, $\bmB$ and $\bmU$, and fluctuating parts, $\bb$ and $\bu$.
Starting from the induction equation governing $\bB$ it is concluded
that the mean magnetic field $\bmB$ has to obey
\EQ
\p_t \bmB = \eta \bnab^2 \bmB + \bnab \x (\bmU \x \bmB + \bscE ) \, ,
    \quad \bnab \cdot \bmB = 0 \, .
\label{eq01}
\EN
Here, $\eta$ means the magnetic diffusivity of the fluid, for simplicity considered
as independent of position, and $\bscE$ the mean electromotive force caused
by the velocity and magnetic fluctuations,
\EQ
\bscE = \lan \bu \x \bb \rangle \, .
\label{eq03}
\EN
Mean fields are defined by some kind of averaging satisfying the Reynolds rules.
They are denoted either by overbars or synonymously by angle brackets.

The induction equation governing $\bB$ also implies
\EQA
&& \!\!\!\!\!\!\!\!\!\!\!
    \p_t \bb = \eta \bnab^2 \bb + \bnab \x [ (\bu \x \bb)' + \bmU \x \bb + \bu \x \bmB],
\nonumber\\
&& \qquad \qquad \qquad \qquad \qquad
    \bnab \cdot \bb = 0 \, ,
\label{eq05}
\ENA
where $(\bu \x \bb)' = \bu \x \bb - \lan \bu \x \bb \ran$.
With this in mind we may conclude that $\bscE$
can be represented as a sum of two parts,
\EQ
\bscE = \bscE^{(0)} + \bscE^{(\mB)} \, ,
\label{eq07}
\EN
where $\bscE^{(0)}$ is a functional of $\bu$ and $\bmU$,
and $\bscE^{(\mB)}$ a functional of $\bu$, $\bmU$ and $\bmB$,
which is linear in $\bmB$ but vanishes if $\bmB$
is zero everywhere and at all past times
(see, e.g., R\"adler 1976, 2000, R\"adler \& Rheinhardt 2007).
These statements apply independently on whether or not $\bu$ or $\bmU$ depend on $\bmB$.
If they depend on $\bmB$ and the total variation of $\bscE$ with $\bmB$ is considered,
$\bscE^{(0)}$ may well depend on $\bmB$,
and $\bscE^{(\mB)}$ need not be linear in $\bmB$.

A non--zero $\bscE^{(0)}$ corresponds to a battery.
Assume for a moment that equation (\ref{eq01}) for $\bmB$ with $\bscE = \bzo$
has no growing solutions.
If then $\bscE^{(0)}$ takes non--zero values, but $\bscE^{(\mB)}$ remains equal to zero,
$\bmB$ grows, even if initially equal to zero, to a finite magnitude determined by $\bscE^{(0)}$.
If, on the other hand, $\bscE^{(0)}$ remains equal to zero a non--zero $\bscE^{(\mB)}$ may allow
(if it has a suitable structure) a dynamo, that is, let an arbitrarily small seed magnetic field $\bmB$
grow exponentially (in the absence of back--reaction on the fluid motion even endlessly).
A small non--zero $\bscE^{(0)}$ may deliver a seed field for such a dynamo.
This possibility has been already discussed in the context of young galaxies
(Brandenburg \& Urpin 1998).

In most of the general representations and applications of mean--field magnetohydrodynamics
the part $\bscE^{(0)}$ of the electromotive force $\bscE$ has been ignored.
Indeed, if it occurs at all, it decays to zero in the course of time
except in cases in which an independent magnetohydrodynamic turbulence exists,
e.g., as a result of a small--scale dynamo.

The possibility of a non--zero $\bscE^{(0)}$ due to local, that is small--scale, dynamos in the solar convection zone
has been discussed by R\"adler (1976).
We express his statements here by
\EQ
\bscE^{(0)} = c_\gamma \bgamma + c_\Omega \bOmega + c_{\gamma \Omega} \bgamma \x \bOmega \, ,
\label{eq09}
\EN
where $\bgamma$ is a gradient, e.g., of the turbulence intensity,
$\bOmega$ the angular velocity responsible for the Coriolis force,
and $c_\gamma$, $c_\Omega$ and $c_{\gamma \Omega}$ some coefficients.
More precisely, $c_\gamma$ and $c_{\gamma \Omega}$ are scalars and $c_\Omega$ is a pseudoscalar.

Another interesting result has been derived by Yoshizawa (1990,
see also Yoshizawa 1993 or  Yoshizawa, Itoh \& Itoh 2003).
Considering an originally homogeneous isotropic magnetohydrodynamic turbulence under the influence of a mean flow
or a rigid--body rotation, or both, he found
\EQ
\bscE^{(0)} = c_W \bW + c_\Omega \bOmega \, ,
\label{eq11}
\EN
where $\bW = \bnab \x \bmU$ is the mean vorticity,
$\bOmega$ again the angular velocity responsible for the Coriolis force,
and $c_W$ and $c_\Omega$ are pseudoscalars coefficients which are, roughly speaking,
proportional to the cross--helicity $\lan \bu \cdot \bb \ran$.
This result has recently been used for an interpretation of the Archontis dynamo
(Sur \& Brandenburg 2009).

The main purpose of this paper is to demonstrate that a mean electromotive force $\bscE^{(0)}$ proportional
to the mean fluid velocity $\bmU$ may occur in originally homogeneous isotropic magnetohydrodynamic turbulence.
This should be expected as soon as there is a non--zero correlation between the fluctuating parts
of velocity and electric current, $\bu$ and $\bj = \mu_0^{-1} \bnab \x \bb$, or, what is equivalent,
between the fluctuating parts of vorticity and magnetic field, $\bomega = \bnab \x \bu$ and $\bb$;
as usual, $\mu_0$ means the magnetic permeability.
We express this condition roughly by saying that $\lan \bu \cdot \bj \ran$ or $\lan \bomega \cdot \bb \ran$
have to be unequal to zero.
Unlike $\lan \bu \cdot \bb \ran$, which characterizes the linkage between vortex tubes and magnetic flux tubes,
$\lan \bu \cdot \bj \ran$ quantifies the linkage between vortex tubes and current tubes.

In Section \ref{sec2} we explain the basis of our calculations and provide general relations
for the determination of the mean electromotive force $\bscE^{(0)}$.
In Section \ref{sec3} we derive results for homogeneous isotropic turbulence,
in particular the last--mentioned one, and we also reproduce that given by (\ref{eq11}).
Proceeding then in Section \ref{sec4} to inhomogeneous turbulence we report on results
related to those indicated in (\ref{eq09}).
The relevance of the results obtained in this paper
and the need of further work are discussed in Section \ref{sec5}.

\section{General concept}
\label{sec2}

\subsection{Basic equations}

We consider a magnetic field $\bB$ in a homogeneous incompressible electrically conducting turbulent fluid
in a rotating frame.
It is assumed that $\bB$ and the fluid velocity $\bU$ are governed by
\EQA
&& \!\!\!\!\!\!\!\!
    \p_t \bB = \eta \bnab^2 \bB + \bnab \x (\bU \x \bB + \bH) \, , \quad \bnab \cdot \bB = 0
\nonumber\\
&& \!\!\!\!\!\!\!\!
    \p_t \bU + \bU \cdot \bnab \bU = - \varrho^{-1} \bnab P + \nu \bnab^2 \bU
\label{eq21}\\
&& \qquad \qquad \qquad \qquad \!\!\!
    - 2 \bOmega \x \bU + \bF \, , \quad \,  \bnab \cdot \bU = 0 \, ,
\nonumber
\ENA
where $\eta$ is again the magnetic diffusivity, $\nu$ the kinematic viscosity
and $\varrho$ the mass density of the fluid.
$P$ is the sum of hydrostatic and centrifugal pressure
and $\bOmega $ the angular velocity defining the Coriolis force.
The external electromotive force $\bH$ and the external ponderomotive force $\bF$
will allow us to mimic magnetohydrodynamic turbulence.
For the sake of simplicity we have ignored the back--reaction of the magnetic field
on the fluid motion.

Adopting the mean--field concept and taking averages of (\ref{eq21}) we arrive
at equations for the mean fields $\bmB$ and $\bmU$.
The equation for $\bmB$ differs from (\ref{eq01}) only in so far as instead of $\bmU \x \bmB$ the sum $\bmU \x \bmB + \bmH$ occurs.
For the magnetic and velocity fluctuations $\bb$ and $\bu$ we further may derive
\EQA
&& \!\!\!\!\!\!\!\!
    \p_t \bb = \eta \bnab^2 \bb + \bnab \x \left[(\bu \x \bb)' + \bmU \x \bb + \bu \x \bmB + \bh\right]
\nonumber\\
&& \qquad \qquad \qquad \qquad \qquad \qquad
    \bnab \cdot \bb = 0
\nonumber\\
&& \!\!\!\!\!\!\!\!
    \p_t \bu = - \varrho^{-1} \bnab p + \nu \bnab^2 \bu - (\bu \cdot \bnab \bu)'
\label{eq23}\\
&& \qquad \qquad \;\;
    - 2 \bOmega \x \bu - \bmU \cdot \bnab \bu - \bu \cdot \bnab \bmU + \bff
\nonumber\\
&& \qquad \qquad \qquad \qquad \qquad \qquad
    \bnab \cdot \bu = 0 \, ,
\nonumber
\ENA
where again $(\bu \x \bb)'$ stands for $\bu \x \bb - \lan \bu \x \bb \ran$,
analogously $(\bu \cdot \bnab \bu)'$ for $\bu \cdot \bnab \bu - \lan \bu \cdot \bnab \bu \ran$,
and $\bh$, $p$ and $\bff$ are the fluctuating parts of $\bH$, $P$ and $\bF$.
The equations for $\bb$ differ from (\ref{eq05}) only by the additional electromotive force $\bh$.

We strive to calculate the part $\bscE^{(0)}$ of the mean electromotive force.
So we put simply $\bmB = \bzo$ in (\ref{eq23}).
This is not germane to the following considerations and could always be justified by choosing a suitable $\bmH$.
Basically the remainder of equations (\ref{eq23}) with a given $\bmU$ implies the possibility
of a small--scale dynamo, that is of non--decaying $\bb$, even if $\bh$ is equal to zero.
In what follows we introduce however some further simplifying assumptions which undermine this possibility,
and we mimic a small--scale dynamo with a proper non--zero $\bh$.

Let us assume that $\bu$ and $\bb$ depend only weakly on $\bmU$ and $\bOmega$
so that $\bscE^{(0)}$ is linear in these quantities.
We further assume that $\bmU$ varies only weakly in space and time
so that $\bscE^{(0)}$ in a given point depend only on $\bmU$ and its first spatial derivatives in this point.
Thus we have
\EQ
{\cE}^{(0)}_i = {\cE}^{(00)}_i + a_{ip} \mU_p  + b_{ipq} \p \mU_p / \p x_q + c_{ip} \Omega_p \, ,
\label{eq25}
\EN
where ${\cE}^{(00)}_i$ as well as the coefficients $a_{ip}$, $b_{ipq}$ and $c_{ip}$
are independent of $\bmU$ and $\bOmega$.

We now split $\bb$ and $\bu$ according to
\EQ
\bb = \bb^{(0)} + \bb^{(1)} + \cdots \, , \quad \bu = \bu^{(0)} + \bu^{(1)} + \cdots
\label{eq27}
\EN
into parts $\bb^{(0)}$ and $\bu^{(0)}$ independent of $\bmU$ and $\bOmega$,
parts $\bb^{(1)}$ and $\bu^{(1)}$ of first order in $\bmU$ or $\bOmega$
and higher-order contributions, which are however not considered in what follows.
The assumption of the linearity of $\bscE^{(0)}$ in $\bmU$ and $\bOmega$ implies
\EQ
\bscE^{(0)} = \lan \bu^{(0)} \x \bb^{(0)} \ran + \lan \bu^{(0)} \x \bb^{(1)} \ran
    + \lan \bu^{(1)} \x \bb^{(0)} \ran \, .
\label{eq29}
\EN

Returning to equations (\ref{eq23}) and considering $\bh$ and $\bff$ as independent of $\bmU$ and $\bOmega$,
we find for $\bb^{(0)}$ and $\bu^{(0)}$
\EQA
&& \!\!\!\!\!\!\!\!
\p_t \bb^{(0)} = \eta \bnab^2 \bb^{(0)} + \bnab \x [(\bu^{(0)} \x \bb^{(0)})' + \bh]
\nonumber\\
&& \qquad \qquad \qquad \qquad \qquad \qquad
    \bnab \cdot \bb^{(0)} = 0
\nonumber\\
&& \!\!\!\!\!\!\!\!
\p_t \bu^{(0)} = - \varrho^{-1} \bnab p^{(0)} + \nu \bnab^2 \bu^{(0)}
\label{eq31}\\
&& \qquad \qquad \qquad \qquad
  - (\bu^{(0)} \cdot \bnab \bu^{(0)})' + \bff
\nonumber\\
&& \qquad \qquad \qquad \qquad \qquad \qquad
   \bnab \cdot \bu^{(0)} = 0 \, .
\nonumber
\ENA
In the following we denote the turbulence defined by $\bb^{(0)}$ and $\bu^{(0)}$
as ``background turbulence".
In the equations resulting for $\bb^{(1)}$ and $\bu^{(1)}$ we introduce
some generalized second--order correlation approximation,
that is, neglect all terms originating from $(\bu \x \bb)'$ and $(\bu \cdot \bnab \bu)'$.
Hence we have
\EQA
&& \!\!\!\!\!\!\!\!
\p_t \bb^{(1)} = \eta \bnab^2 \bb^{(1)} + \bnab \x (\bmU \x \bb^{(0)})
\nonumber\\
&& \qquad \qquad \qquad \qquad \qquad
    \bnab \cdot \bb^{(1)} = 0
\nonumber\\
&& \!\!\!\!\!\!\!\!
\p_t \bu^{(1)} = - \frac{1}{\varrho} \bnab p^{(1)} + \nu \bnab^2 \bu^{(1)}
\label{eq33}\\
&& \qquad
    - 2 \bOmega \x \bu^{(0)} - \bmU \cdot \bnab \bu^{(0)} - \bu^{(0)} \cdot \bnab \bmU
\nonumber\\
&& \qquad \qquad \qquad \qquad \qquad
    \bnab \cdot \bu^{(1)} = 0 \, .
\nonumber
\ENA

\subsection{Relation for $\bscE^{(0)}$}

In the following derivations we use a Fourier transformation of the form
\EQ
F(\bx, t) = \int \!\!\! \int \hat {F} (\bk, \omega) \,
    \exp [\iu (\bk \cdot \bx - \omega t)] \, \dd^3 k \, \dd \omega \, .
\label{eq41}
\EN
The integrations are over all $\bk$ and $\omega$.

In view of the determination of $\bscE^{(0)}$ we first note
\EQ
\lan \bu (\bx, t) \x \bb (\bx, t) \ran_i =
    \epsilon_{ijk} \!\! \int \!\!\! \int \hat{Q}_{jk} (\bx, t; \bk, \omega) \, \dd^3 k \, \dd \omega \, ,
\label{eq43}
\EN
where $\hat{Q}_{jk} (\bx, t; \bk, \omega)$ is the Fourier transform of
\EQA
&& \!\!\!\!\!\!\!\!
    Q_{jk} (\bx, t; \bxi, \tau) =
\nonumber\\
&& \!\!\!\!\!\!\!\!
    \lan u_j (\bx + \bxi/2, t + \tau/2) \, b_k (\bx - \bxi/2, t - \tau/2) \ran
\label{eq45}
\ENA
with respect to $\bxi$ and $\tau$.
Adopting the formalism of Roberts \& Soward (1975) we find that
\EQA
&& \!\!\!\!\!\!\!\!\!\!\!\!\!\!
    \hat{Q}_{jk} (\bx, t; \bk, \omega) =
\nonumber\\
&& \!\!\!\!\!\!\!\!\!\!\!\!\!\!
    \int \!\!\! \int \lan \hu_j (\bk + \bk'/2, \omega + \omega'/2) \,
    \hb_k (-\bk + \bk'/2, -\omega + \omega'/2) \ran
\nonumber\\
&& \qquad \qquad \qquad
    \exp[\iu (\bk' \cdot \bx - \omega' t)] \, \dd^3 k' \, \dd \omega' \, ;
\label{eq47}
\ENA
see Appendix \ref{robsow}.
As a consequence of $\bnab \cdot \bu = \bnab  \cdot \bb = 0$ the conditions
\EQ
\nabla_j \hat{Q}_{jk} + 2 \, \iu k_j \hat{Q}_{jk} = 0 \, , \quad
    \nabla_k \hat{Q}_{jk} - 2 \, \iu k_k \hat{Q}_{jk} = 0
\label{eq49}
\EN
have to be satisfied.
Note that for the determination of $\bscE^{(0)}$ only the antisymmetric part of $\hat{Q}_{jk}$ is needed.

Considering $\bscE^{(0)}$ we restrict our attention simply to $\bx = \bzo$.
In that sense we put
\EQ
\mU_i = U_i + U_{ij} x_j \, .
\label{eq53}
\EN
We consider the equations (\ref{eq31}) as solved, that is, $\bb^{(0)}$ and $\bu^{(0)}$
as known.
Subjecting then the equations (\ref{eq33}) for $\bb^{(1)}$ and $\bu^{(1)}$ with $\bmU$ specified according to (\ref{eq53})
to a Fourier transformation and eliminating the pressure term in the usual way we find
\EQA
&& \!\!\!\!\!\!\!\!
    \hb^{(1)}_i = - E \big[ \iu k_m U_m \hb^{(0)}_i - U_{im} \hb^{(0)}_m
    - k_m U_{mn} \p \hb^{(0)}_i / \p k_n \big]
\nonumber\\
&& \qquad \qquad
   E = (\eta k^2 - \iu \omega)^{-1} \, , \quad \hb_i k_i = 0
\nonumber\\
&& \!\!\!\!\!\!\!\!
    \hu^{(1)}_i = - N \big[ \iu k_m U_m \hu^{(0)}_i + U_{im} \hu^{(0)}_m
\nonumber\\
&& \qquad \qquad
    - k_m U_{mn} (2 k_i \hu^{(0)}_n / k^2 + \p \hu^{(0)}_i / \p k_n)
\label{eq55}\\
&& \qquad \qquad
    + 2 \epsilon_{imn} k_m (\bk \cdot \bOmega) \hu^{(0)}_n / k^2 \big]
\nonumber\\
&& \qquad \qquad
    N = (\nu k^2 - \iu \omega)^{-1} \, , \quad  \hu_i k_i = 0 \, .
\nonumber
\ENA

Calculating now $\hat{Q}_{jk}$ on the basis of (\ref{eq47}) and (\ref{eq55}) we neglect again all contributions
of higher than first order in $\bmU$ and $\bOmega$.
We further discard terms with more than one spatial derivative,
in particular products of $U_{ij}$ with any other spatial derivative.
Since $\hat{Q}_{jk}$ should only weakly vary with $\bx$ we expand $\lan \hu_j \hb_k \ran$ in (\ref{eq47})
for small $\bk'$ and arrive so at
\EQA
&& \!\!\!\!\!
    \hat{Q}_{jk} = \hat{Q}^{(0)}_{jk} + \iu (E^* - N) \, (\bk \cdot \bU) \, \hat{Q}^{(0)}_{jk}
\nonumber\\
&& + E^* U_{km} \, \hat{Q}^{(0)}_{jm}  - N U_{jm} \, \hat{Q}^{(0)}_{mk}
    + 2 N U_{mn} k_j k_m \hat{Q}^{(0)}_{nk}
\nonumber\\
&& + {\textstyle{1 \over 2}} ({E^*}' + N') U_{mn} k_m k_n \hat{Q}^{(0)}_{jk} / k^2
\nonumber\\
&& + {\textstyle{1 \over 2}} (E^* + N) U_{mn} k_m \p \hat{Q}^{(0)}_{jk} / \p k_n
\nonumber\\
&& - 2 N \, \epsilon_{jmn} k_m (\bk \cdot \bOmega) \hat{Q}^{(0)}_{nk} / k^2
\nonumber\\
&& - {\textstyle{1 \over 2}} ({E^*} + N) \, (\bU \cdot \bnab) \hat{Q}^{(0)}_{jk}
\label{eq61}\\
&& - {\textstyle{1 \over 2}} ({E^*}' + N') \, (\bk \cdot \bU) \, (\bk \cdot \bnab) \hat{Q}^{(0)}_{jk} /k^2
\nonumber\\
&& + \iu \epsilon_{jmn} \big[ N ((\bk \cdot \bOmega) \nabla_m + k_m (\bOmega \cdot \bnab)) \hat{Q}^{(0)}_{nk} / k^2
\nonumber\\
&& \qquad \quad
    - (2 N - N') k_m (\bk \cdot \bOmega) (\bk \cdot \bnab) \hat{Q}^{(0)}_{nk} / k^4 \big] \, .
\nonumber
\ENA
Here $\hat{Q}^{(0)}_{jk}$ means $\hat{Q}_{jk}$ for the background turbulence,
that is, with $\hat{\bu}$ and $\hat{\bb}$ replaced by $\hat{\bu}^{(0)}$ and $\hat{\bb}^{(0)}$.
For simplicity arguments are dropped; those of $\hat{Q}_{jk}$ and $\hat{Q}^{(0)}_{jk}$ are $(\bx, t; \bk, \omega)$,
those of $E$, $N$, etc. are $(k,\omega)$.
The asterisk means complex conjugation and $F' = k \p F / \p k$.

Returning now to the representation (\ref{eq25}) of $\bscE^{(0)}$ we find
\EQA
&& \!\!\!\!\!\!\!\!
    {\cE}^{(00)}_i = \epsilon_{ijk} \int \!\!\!\! \int \hat{Q}^{(0)}_{jk} \, \dd^3 k \dd \omega
\nonumber\\
&& \!\!\!\!\!\!\!\!
    a_{ip} = \epsilon_{ijk} \int \!\!\!\! \int \big[ \iu (E^* - N) k_p
    - {\textstyle{1 \over 2}} (E^* + N) \nabla_p
\nonumber\\
&& \quad
- {\textstyle{1 \over 2}} ({E^*}' + N') (k_p / k^2) (\bk \cdot \bnab) \big]
    \hat{Q}^{(0)}_{jk} \dd^3k \, \dd \omega
\nonumber\\
&& \!\!\!\!\!\!\!\!
    b_{ipq} = \epsilon_{ijk} \int \!\!\!\! \int
    \big[ E^* \delta_{kp} \hat{Q}^{(0)}_{jq}
    + {\textstyle{1 \over 2}}(E^* + N)
    k_p \p \hat{Q}_{jk} / \p k_q
\nonumber\\
&& \quad
   - N (\delta_{jp} - k_j k_p / k^2) \hat{Q}^{(0)}_{qk} \big] \, \dd^3 k \, \dd \omega
\label{eq67}\\
&& \!\!\!\!\!\!\!\!
    c_{ip} = \int \!\!\!\! \int \big[ (N/k^2) \big( 2 k_i k_p - \iu (k_i \nabla_p + k_p \nabla_i) \big)
    \hat{Q}^{(0)}_{ll}
\nonumber\\
&& \quad
    + \iu (N/k^2) (k_p \nabla_k + k_k \nabla_p) \hat{Q}^{(0)}_{ik}
\nonumber\\
&& \quad
    - \iu (2 N - N') (k_p/k^4) (\bk \cdot \bnab)
\nonumber\\
&& \qquad \qquad \qquad \quad
    \big(k_k \hat{Q}^{(0)}_{ik} - k_i \hat{Q}^{(0)}_{ll} \big) \big] \dd^3 k \, \dd \omega \, ,
\nonumber
\ENA
where again the above remarks on arguments apply.

\section{Homogeneous isotropic turbulence}
\label{sec3}

\subsection{General result}

Consider now the simple case in which the background turbulence is homogeneous and isotropic
and return first to (\ref{eq25}).
Since there is no isotropic vector we have ${\cE}^{(00)}_i = 0$.
Isotropy further implies
$a_{ip} = c_U \delta_{ip}$, $b_{ijk} = c_W \epsilon_{ijk}$ and $c_{ip} = c_\Omega \delta_{ij}$.
Hence we obtain
\EQ
\bscE^{(0)} = c_U \, \bmU + c_W \, \bnab \x \bmU + c_\Omega \, \bOmega
\label{eq73}
\EN
with a scalar $c_U$ and pseudoscalars $c_W$ and $c_\Omega$.

Due to homogeneity and isotropy of the turbulence we have
\EQA
&& \!\!\!\!\!\!\!\!\!\!\!\!
    \hat{Q}^{(0)}_{jk} (\bk, \omega) =
\nonumber\\
&& \!\!\!\!\!\!\!\!\!\!\!\!
    \frac{1}{2} \Big[ \big( \delta_{jk} - \frac{k_j k_k}{k^2} \big) \hat{\Phi}^{(0)} (k, \omega)
    - \iu \epsilon_{jkl} \frac{k_l}{k^2} \hat{\Psi}^{(0)} (k, \omega) \Big] \, ,
\label{eq75}
\ENA
where $\hat{\Phi}^{(0)}$ and $\hat{\Psi}^{(0)}$ are the Fourier transforms of
\EQA
\Phi^{(0)} &=&
\nonumber\\
&& \!\!\!\!\!\!\!\!\!\!\!\!\!\!\!\!\!\!\!\!\!\!\!
    \lan \bu^{(0)} (\bx + \bxi/2, t + \tau/2) \cdot \bb^{(0)} (\bx - \bxi/2, t - \tau/2) \ran
\nonumber\\
\Psi^{(0)} &=&
\label{eq77}\\
&& \!\!\!\!\!\!\!\!\!\!\!\!\!\!\!\!\!\!\!\!\!\!\!
    \mu_0 \lan \bu^{(0)} (\bx+ \bxi/2, t + \tau/2) \cdot \bj^{(0)} (\bx - \bxi/2, t - \tau/2) \ran
\nonumber
\ENA
with respect to $\bxi$ and $\tau$,
and $\mu_0 \bj^{(0)}$ stands for $\bnab \x \bb^{(0)}$.
The homogeneity implies that $\Phi^{(0)}$ and $\Psi^{(0)}$ as well as $\hat{\Phi}^{(0)}$ and $\hat{\Psi}^{(0)}$
are independent of $\bx$ and further that
\EQA
&& \!\!\!\!\!\!\!\!\!\!\!\!
    \mu_0 \lan \bu^{(0)} (\bx+ \bxi/2, t + \tau/2) \cdot \bj^{(0)} (\bx - \bxi/2, t - \tau/2) \ran =
\nonumber\\
&& \!\!\!\!\!\!\!\!\!\!\!\!
    \lan \bomega^{(0)} (\bx+ \bxi/2, t + \tau/2) \cdot \bb^{(0)} (\bx - \bxi/2, t - \tau/2) \ran \, ,
\label{eq79}
\ENA
where $\bomega^{(0)}$ is the vorticity of the velocity field $\bu^{(0)}$, that is $\bomega^{(0)} = \bnab \x \bu^{(0)}$.
In general $\hat{Q}_{jk}$, $\Phi^{(0)}$, $\Psi^{(0)}$ as well as $\hat{\Phi}^{(0)}$ and $\hat{\Psi}^{(0)}$ may depend on $t$.
If we however assume that the turbulence shows in addition to its homogeneity also statistical steadiness
this dependence vanishes.
In addition the arguments $(\bx + \bxi/2, t + \tau/2)$ and $(\bx - \bxi/2, t - \tau/2)$ in (\ref{eq77}) and (\ref{eq79})
may then be replaced, e.g., by $(\bx, t)$ and $(\bx - \bxi, t - \tau)$ or by $(\bx + \bxi, t + \tau)$ and $(\bx, t)$,
respectively.

With (\ref{eq67}) and (\ref{eq75}) we find
\EQA
c_U &=& - \frac{2}{3} \int \!\!\!\! \int ( E^* - N ) \, \hat{\Psi}^{(0)} \,  k \, \dd^3k \, \dd \omega
\nonumber\\
c_W &=& \frac{1}{3} \int \!\!\!\! \int ( 2 E^* + N ) \, \hat{\Phi}^{(0)} \, \dd^3k \, \dd \omega
\label{eq81}\\
c_\Omega &=& \frac{4}{3} \int \!\!\!\! \int N \, \hat{\Phi}^{(0)} \, \dd^3k \, \dd \omega \, .
\nonumber
\ENA
We point out that $E = (2 \pi)^4 \hat{G^{(\eta)}}$ and $N = (2 \pi)^4 \hat{G^{(\nu)}}$
where the $G^{(\gamma)}$ are the well--known Green's functions defined by
\EQA
G^{(\gamma)} (\xi, \tau) &=& (4 \pi \gamma \tau)^{-3/2} \exp (- \xi^2 / 4 \gamma \tau ) \;\; \mbox{for} \; \tau > 0
\nonumber\\
G^{(\gamma)} (\xi, \tau) &=& 0  \;\; \mbox{for} \; \tau \leq 0 \, .
\label{eq83}
\ENA
Considering this and applying the convolution theorem to (\ref{eq81}) we obtain
\EQA
&& \!\!\!\!\!\!\!\!
    c_U = - \frac{\mu_0}{3} \int \!\!\!\! \int \big(G^{(\eta)} (\xi, \tau) - G^{(\nu)} (\xi, \tau) \big)
\nonumber\\
&& \qquad
    \lan \bu^{(0)} (\bx, t) \cdot \bj^{(0)} (\bx - \bxi, t - \tau) \ran \, \dd^3 \xi \, \dd \tau
\nonumber\\
&& \!\!\!\!\!\!\!\!
    c_W = \frac{1}{3} \int \!\!\!\! \int \big(G^{(\eta)} (\xi, \tau) + \frac{1}{2} G^{(\nu)} (\xi, \tau) \big)
\label{eq87}\\
&& \qquad
    \lan \bu^{(0)} (\bx, t) \cdot \bb^{(0)} (\bx - \bxi, t - \tau) \ran \,
    \dd^3 \xi \, \dd \tau
\nonumber\\
&& \!\!\!\!\!\!\!\!
    c_\Omega = - \frac{2}{3} \int \!\!\!\! \int G^{(\nu)} (\xi, \tau)
\nonumber\\
&& \qquad
    \lan \bu^{(0)} (\bx, t) \cdot \bb^{(0)} (\bx - \bxi, t - \tau) \ran \, \dd^3 \xi \, \dd \tau \, .
\nonumber
\ENA
Here the integrations are over all $\bxi$ and primarily also over all $\tau$.
However, since the $G^{(\eta)} = G^{(\nu)} = 0$ for $\tau \leq 0$, they involve in fact only positive $\tau$.

The most remarkable result of our derivations is that a contribution to $\bscE^{(0)}$ proportional to $\bmU$,
that is, a term $c_U \bmU$ in (\ref{eq73}), may occur.
This possibility has not previously been considered in the literature.
According to (\ref{eq87}), as long as $\lan \bu^{(0)} (\bx, t) \cdot \bj^{(0)} (\bx - \bxi, t - \tau) \ran$
does not vanish and $\eta \not= \nu$, the coefficient $c_U$ may well be different from zero.
In the special case $\eta = \nu$, however, it is equal to zero.
In what follows the occurrence of that contribution to $\bscE^{(0)}$ proportional to $\bmU$
is labeled as ``$\lan \bu \cdot \bj \ran$ effect".

For non--vanishing $\lan \bu^{(0)} (\bx, t) \cdot \bb^{(0)} (\bx - \bxi, t - \tau) \ran$
the pseudoscalar coefficients $c_W$ and $c_\Omega$ will in general be different from zero.
Then, as already found by Yoshizawa (1990), contributions
to $\bscE^{(0)}$
proportional to the mean vorticity \mbox{$\bnab \x \bmU$} and to the angular velocity $\bOmega$ will occur.
We refer to them as ``$\lan \bu \cdot \bb \ran$ effects" or ``Yoshizawa effects".

We stress that the $\lan \bu \cdot \bj \ran$ effect is well possible under circumstances
in which $\lan \bu^{(0)} (\bx, t) \cdot \bb^{(0)} (\bx - \bxi, t - \tau) \ran$ is equal to zero,
that is, in which there are no
$\lan \bu \cdot \bb \ran$ effects.

\subsection{Special cases}

Let us first consider $\bscE^{(0)}$ in some limiting cases with respect to $\eta$ and $\nu$.
We use the fact that
\EQ
G^{(\gamma)} (\xi, \tau) \to \delta^3 (\bxi) \quad \mbox{as} \;\; \gamma \to 0 \, .
\label{eq89}
\EN
In the limit defined by $\eta \to 0$ and $\nu \to \infty$ we obtain
\EQ
c_U = - \frac{1}{3} A \, , \quad
    c_W = \frac{1}{3} C \, , \quad
    c_\Omega = 0 \, ,
\nonumber
\label{eq91}
\EN
in the limit $\eta \to \infty$ and $\nu \to 0$
\EQ
c_U = \frac{1}{3} A \, , \quad
    c_W = \frac{1}{6} C \, , \quad
    c_\Omega = - \frac{2}{3} C \, ,
\label{eq93}
\EN
and in the limit $\eta, \, \nu \to 0$
\EQ
c_W = 0 \, , \quad
    c_W = \frac{1}{2} C \, , \quad
    c_\Omega = - \frac{2}{3} C \, ,
\label{eq95}
\EN
where
\EQA
A &=& \mu_0 \int_0^\infty \lan \bu^{(0)} (\bx, t)
    \cdot \bj^{(0)} (\bx, t - \tau) \ran \, \dd \tau
\nonumber\\
C &=& \int_0^\infty \lan \bu^{(0)} (\bx, t)
    \cdot \bb^{(0)} (\bx, t - \tau)) \ran \, \dd \tau \, .
\label{eq97}
\ENA
Instead of the last relations we may also write
\EQA
A &=& \mu_0 \lan \bu^{(0)} (\bx, t) \cdot \bj^{(0)} (\bx, t) \ran \, \tau_A
\label{eq99}\\
C &=& \lan \bu^{(0)} (\bx, t) \cdot \bb^{(0)} (\bx, t)) \ran \, \tau_C
\nonumber
\ENA
with correlation times $\tau_A$ and $\tau_C$ defined by equating the respective right--hand sides
of (\ref{eq97}) and (\ref{eq99}).

In view of a numerical test we also consider the case in which
$\lan \bu^{(0)} (\bx, t) \cdot \bj^{(0)} (\bx - \bxi, t - \tau) \ran$
and $\lan \bu^{(0)} (\bx, t) \cdot \bb^{(0)} (\bx - \bxi, t - \tau) \ran$,
as far as they enter into the integrals in (\ref{eq87}), do not markedly vary with $\tau$.
Ignoring the dependence on $\tau$ completely and using
\begin{equation}
\int_0^\infty G^{(\gamma)} (\xi, \tau) \, \dd \tau = \frac{1}{4 \pi \gamma \xi}
\label{eq101}
\end{equation}
we find
\EQA
c_U &=& - \frac{1}{12 \pi}\left(\frac{1}{\eta} - \frac{1}{\nu}\right) \, A^\dag
\nonumber\\
c_W &=& \frac{1}{12 \pi} \left(\frac{1}{\eta} + \frac{1}{2 \nu}\right) \, C^\dag \, , \quad
c_\Omega = - \frac{1}{6 \pi \nu} \, C^\dag
\label{eq103}
\ENA
with
\EQA
A^\dag &=& \int_\infty \lan \bu^{(0)} (\bx, t) \cdot
    (\bnab \x \bb^{(0)} (\bx - \bxi, t)) \ran \, \frac{\dd^3 \xi}{\xi}
\nonumber\\
C^\dag &=& \int_\infty \lan \bu^{(0)} (\bx, t)
    \cdot \bb^{(0)} (\bx - \bxi, t) \ran \, \frac{\dd^3 \xi}{\xi} \, .
\label{eq105}
\ENA
We may introduce vector potentials $\bpsi^{(0)}$ and $\ba^{(0)}$ such that
\begin{eqnarray}
\bnab \x \bpsi^{(0)} &=& \bu^{(0)} \, , \quad \bnab \cdot \bpsi^{(0)} = 0
\nonumber\\
\bnab \x \ba^{(0)} &=& \bb^{(0)} \, , \quad \bnab \cdot \ba^{(0)} = 0
\label{eq105}
\end{eqnarray}
and therefore
\begin{eqnarray}
\bpsi^{(0)} (\bx) &=& \frac{1}{4 \pi} \int_\infty \bnab \x \bu^{(0)} (\bx - \bxi) \, \frac{\dd^3 \xi}{\xi}
\nonumber\\
\ba^{(0)} (\bx) &=& \frac{1}{4 \pi} \int_\infty \bnab \x \bb^{(0)} (\bx - \bxi) \, \frac{\dd^3 \xi}{\xi} \, .
\label{eq107}
\end{eqnarray}
For simplicity the argument $t$ is omitted everywhere.
The result (\ref{eq103}) can then be written in the form
\EQA
c_U &=& - \frac{1}{3}\left(\frac{1}{\eta} - \frac{1}{\nu}\right) \, A^\ddag
\nonumber\\
c_W &=& \frac{1}{3} \left(\frac{1}{\eta} + \frac{1}{2 \nu}\right) \, C^\ddag \, , \quad
    c_\Omega = - \frac{3}{2 \nu} \, C^\ddag
\label{eq109}
\ENA
with
\EQA
A^\ddag &=& \lan \bu^{(0)} \cdot \ba^{(0)} \ran = \lan \bpsi^{(0)} \cdot \bb^{(0)} \ran
\nonumber\\
C^\ddag &=& \lan \bpsi^{(0)} \cdot \bb^{(0)} \ran = \lan (\bnab \x \bu^{(0)})\cdot \ba^{(0)} \ran \, .
\label{eq111}
\ENA
The arguments of the quantities in the angle brackets are, of course, always $(\bx, t)$.

\subsection{A numerical test}

As a check of the above derivations, the electromotive force $\bscE^{(0)}$ and so the coefficient $c_U$
have been determined with numerical solutions of the equations (\ref{eq23}).
For these calculations $\bmB$ has been put equal to zero, and $\bmU$ was specified via the initial condition
to be constant in space and turned out to remain nearly constant in time, too.
The forcing fields $\bh$ and $\bff$ were taken as periodic in the space coordinates $x$, $y$ and $z$, and steady.
More precisely, $\bh$ and $\bff$ differed only by constant factors from
the vector field $\be (k \bx) \equiv (\sin kz, \sin kx, \sin ky)$, with a constant $k$.
The flow which would result from $\bff$ is the no-cosine ABC flow
of Archontis (2000, see also Dorch \& Archontis 2004 and
Cameron \& Galloway 2006).
Flows of this type are non--helical.
For this reason the $\lan \bu \cdot \bj \ran$ effect should occur,
but no $\lan \bu \cdot \bb \ran$ effects.
Corresponding to the steadiness of $\bh$ and $\bff$ only steady $\bb$ and $\bu$ were considered.
The average which defines mean fields was taken over all $x$, $y$ and $z$ or,
equivalent to this, over a periodic box.
No approximation such as, e.g., the second--order correlation approximation was used.

Let us specify the result for $c_U$ given by (\ref{eq109}) and (\ref{eq111}),
which has been derived in the second--order correlation approximation,
to the described situation.
Relying on (\ref{eq31}) we assume that $\bu^{(0)}$ and $\bb^{(0)}$
are dominated by contributions proportional to $\be (k \bx)$.
So we find
\EQ
c_U = c_0 R_{\rm m} \left(1 - \frac{1}{P_{\rm m}} \right) \, , \quad
   c_0 = - \frac{\mu_0 \lan \bu^{(0)} \cdot \bj^{(0)} \ran}{3 u_{\rm rms}^{(0)} k} \, ,
\label{eq113}
\EN
with the magnetic Reynolds number $R_{\rm m}$ and the magnetic Prandtl number $P_{\rm m}$ defined by
\EQ
R_{\rm m} = u_{\rm rms}^{(0)} / \eta k \, , \quad P_{\rm m} = \nu / \eta \, ,
\label{eq115}
\EN
and $u_{\rm rms}^{(0)} = \lan \bu^{(0) \, 2} \ran^{1/2}$.

\begin{table}
\centering
\caption{Numerically calculated values of $c_U / c_0$ for several $R_{\rm m}$ and $P_{\rm m}$,
to be compared with the values derived in the second--order correlation approximation, $R_{\rm m} (1 - 1/P_{\rm m})$.}
\label{tab1}
\begin{tabular}{cccccc}
\hline\\
$R_{\rm{m}}$ &  $P_{\rm{m}}$ & $c_U / c_0$ & $R_{\rm{m}} (1 - 1/P_{\rm{m}})$ \\
\hline \\
0.2 & 4 &  0.15  &  0.15 \\
    & 2 &  0.10  &  0.10 \\
    & 1 &  $2.6 \cdot 10^{-6}$  &  0 \\
    & 0.2 & $-$0.80  &  $-$0.80 \\
    & 0.04 &  $-$4.80  &  $-$4.80 \\
    & 0.01 &  $-$15.5  &  $-$19.8 \\
\hline\\
1   & 5 &  0.80  &  0.80 \\
    & 1 &  $1.2 \cdot 10^{-6}$  &  0 \\
    & 0.2 &  $-$4.0  &  $-$4.0 \\
 & 0.05 &  $-$17.7  & $-$19.0 \\
\hline\\
10 & 50 &  9.8  & 9.8 \\
    & 10 &  9.0  & 9.0 \\
    & 2 &  4.8  &  5.0 \\
    & 0.5 & $-$9.05  & $-$10.0 \\
\hline\\
\end{tabular}
\end{table}

\begin{figure}[t!]\begin{center}
\includegraphics[width=\columnwidth]{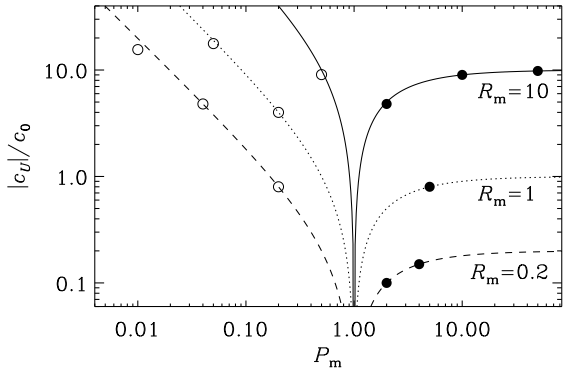}
\end{center}\caption[]{
Numerically obtained values of $|c_U|/c_0$, indicated by $\circ$ for $P_{\rm m} < 1$ (where $c_U < 0$)
but by $\bullet$ for $P_{\rm m} > 1$ (where $c_U > 0$),
and curves representing $R_{\rm m} |1 - 1/P_{\rm m}|$
for $\Rm=10$ (solid), $\Rm=1$ (dotted), and $\Rm=0.2$ (dashed).
}\label{ppm}\end{figure}

In Table \ref{tab1} and Figure~\ref{ppm}
the numerically determined values of $c_U/c_0$ are given in dependence of $R_{\rm m}$ and $P_{\rm m}$.
In agreement with what we have found in our analytical calculations in the second--order correlation approximation
their signs change with growing $P_{\rm m}$ at $P_{\rm m} =1$.
Moreover, in most cases the numerically determined values completely agree with those obtained in this approximation,
that is, with $R_{\rm m} (1 - 1/P_{\rm m})$.
Deviations occur only if the fluid Reynolds number $R_{\rm e} = u_{\rm rms}^{(0)} / \nu k$,
that is $R_{\rm e} = R_{\rm m} / P_{\rm m} $, exceeds a value of about 5.
We should emphasize that all our numerical solutions are laminar and perfectly regular,
just as in Figure~2 (upper row) of Sur \& Brandenburg (2009).

\section{Inhomogeneous turbulence}
\label{sec4}

Let us add some results for the case in which the turbulence is no longer homogeneous
and so also no longer isotropic.
We admit now that quantities like $\lan \bu^{(0)} (\bx + \bxi/2, t + \tau/2) \cdot \bj^{(0)} (\bx - \bxi/2, t - \tau/2) \ran$
and $\lan \bu^{(0)} (\bx + \bxi/2, t + \tau/2) \cdot \bb^{(0)} (\bx - \bxi/2, t - \tau/2) \ran$
may depend on $\bx$ so that their gradients with respect to $\bx$ do not generally vanish.
Then, in addition to the contributions to $\bscE^{(0)}$ given in (\ref{eq73}), other contributions are possible.
Symmetry considerations suggest
\EQA
\bscE^{(0)} &=& c_U \, \bmU + c_W \, \bnab \x \bmU + c_\Omega \, \bOmega
\nonumber\\
&& + \bg + \bg_U \x \bmU + \bg_\Omega \x \bOmega  \, ,
\label{eq121}
\ENA
where $c_U$, $c_W$ and $c_\Omega$ may now vary in space, $\bg$ and $\bg_\Omega$ are gradients of scalars of the type of $c_U$,
and $\bg_U$ is a gradient of a pseudoscalar of type of $c_W$ or $c_\Omega$.
For the sake of simplicity only terms up to the first order in spatial derivatives are regarded here.

We may determine the contributions to $\bscE^{(0)}$ mentioned in (\ref{eq121})
again on the basis of (\ref{eq43}) and (\ref{eq61}).
However, relation (\ref{eq75}) for $\hat{Q}^{(0)}_{ij}$, which applies to homogeneous isotropic turbulence only,
has to be modified.
Admitting now that $\Phi^{(0)}$ and $\Psi^{(0)}$ defined by (\ref{eq77})
and so also $\hat{\Phi}^{(0)}$ and $\hat{\Psi}^{(0)}$ depend on $\bx$
and assuming that their gradients are small we add terms which are linear in these gradients
to the right--hand side of (\ref{eq75}).
Considering further the conditions (\ref{eq49}) we find
\begin{eqnarray}
&& \!\!\!\!\!\!\!\!\!\!\!\!
    \hat{Q}^{(0)}_{jk} (\bx; \bk, \omega) =
\nonumber\\
&& \!\!\!\!\!\!\!\!\!\!\!\!
    \frac{1}{2} \Big[ \big( \delta_{jk} - \frac{k_j k_k}{k^2} \big)
    + \frac{\iu}{2 k^2} \big( k_j \nabla_k - k_k \nabla_j \big) \Big] \,\hat{\Phi}^{(0)} (\bx;  k, \omega)
\nonumber\\
&& \!\!\!\!\!\!\!\!\!\!\!\!
    - \frac{1}{4 k^2} \Big[\epsilon_{jkl} k_l \big( 2 \iu + \frac{1}{k^2} (\bk \cdot \bnab) \big)
\label{eq123}\\
&& \qquad
    - \frac{1}{k^2} (k_j \epsilon_{klm} + k_k \epsilon_{jlm}) k_l \nabla_m \Big] \, \hat{\Psi}^{(0)} (\bx; k, \omega) \, .
\nonumber
\end{eqnarray}
As in the case of homogeneous turbulence $\hat{Q}_{jk}$, $\hat{\Phi}^{(0)}$ and $\hat{\Psi}^{(0)}$
may depend on $t$, what is however of minor importance in this context and therefore not explicitly indicated.

A straightforward calculation confirms then (\ref{eq121}).
As for $c_U$, $c_W$ and $c_\Omega$ we find, as expected, again the relations (\ref{eq81}),
now with $\hat{\Psi}^{(0)}$ and $\hat{\Phi}^{(0)}$ depending in general on $\bx$,
and so also (\ref{eq87}).
Furthermore the calculation yields
\EQA
\bg &=& - \frac{1}{6} \bnab
    \int \!\!\! \int \hat{\Psi}^{(0)} \, k^{-2} \dd^3 k \, \dd \omega
\nonumber\\
\bg_U &=& \frac{1}{3} \bnab \int \!\!\! \int (E^* - N) \, \hat{\Phi}^{(0)} \, \dd^3 k \, \dd \omega
\label{eq125}\\
\bg_\Omega &=& - \frac{2}{3} \bnab
    \int \!\!\! \int N \, \hat{\Psi}^{(0)} \, k^{-2} \, \dd^3 k \, \dd \omega \, .
\nonumber
\ENA
Using again the convolution theorem in combination with above--mentioned connection
between $N$ and $\hat{G^{(\nu)}}$ and between $E$ and $\hat{G^{(\eta)}}$
we arrive at the equivalent relations
\EQA
&& \!\!\!\!\!\!\!\!\!
    \bg = - \frac{\mu_0}{24 \pi} \bnab \int \!\!\! \int \lan \bu^{(0)} (\bx + \bxi/2, t + \tau/2) \cdot
\nonumber\\
&& \qquad \qquad \qquad \qquad
    \bj^{(0)} (\bx - \bxi/2, t - \tau/2) \ran \, \frac{\dd^3 \xi}{\xi} \dd \tau
\nonumber\\
&& \!\!\!\!\!\!\!\!\!
    \bg_U = \frac{1}{6}\bnab \int \!\!\! \int \big( G^{(\eta)} (\bxi, \tau) - G^{(\nu)} (\bxi, \tau) \big)
\label{eq127}\\
&& \!\!\!\!\!\!\!\!\!\!\!\!\!\!\!
    \lan \bu^{(0)} (\bx + \bxi/2, t + \tau/2) \cdot
    \bb^{(0)} (\bx - \bxi/2, t - \tau/2) \ran \, \dd^3 \xi \dd \tau
\nonumber\\
&& \!\!\!\!\!\!\!\!\!
    \bg_\Omega = - \frac{\mu_0}{6 \pi} \bnab \int \!\!\! \int \!\!\! \int
    G^{(\nu)} (\bxi - \bxi', \tau) \frac{\dd^3 \xi'}{\xi'}
\nonumber\\
&& \!\!\!\!\!\!\!\!\!\!\!\!\!\!\!
    \lan \bu^{(0)} (\bx + \bxi/2, t + \tau/2) \cdot
    \bj^{(0)} (\bx - \bxi/2, t - \tau/2) \ran \, \dd^3 \xi \, \dd \tau \, .
\nonumber
\ENA

These results confirm in some sense the statements made in the paper by R\"adler (1976),
formulated above in (\ref{eq09}).
They show however that the vectors $c_\gamma \bgamma$ and $c_{\gamma \Omega} \bgamma$  should not,
as suggested there, be understood in the sense of $\bnab \lan \bu^2 \ran$.
These vectors rather correspond to $\bg$ or $\bg_\Omega$ as given in (\ref{eq125}) and (\ref{eq127}).
Roughly speaking, they should be interpreted in terms of $\bnab \lan \bu \cdot \bj \ran$.

\section{Discussion}
\label{sec5}

The most remarkable result of our calculations is that the mean electromotive force $\bscE^{(0)}$
in a homogeneous isotropic magnetohydrodynamic turbulence
may have a contribution proportional to the mean fluid velocity $\bmU$,
that is, $\bscE^{(0)} = c_U \, \bmU + \cdots$.
We have labeled the occurrence of this contribution as $\lan \bu \cdot \bj \ran$ effect.
The coefficient $c_U$ turned out to be in general unequal to zero
if only a non--zero correlation exists between the fluctuating parts
of the fluid velocity and the electric current density,
$\bu$ and $\bj = \mu_0^{-1} \bnab \x \bb$,
or between the fluctuating parts of the vorticity and the magnetic field,
$\bomega = \bnab \x \bu$ and $\bb$.
As far as the second--order correlation approximation applies, $c_U$ vanishes for $\eta = \nu$,
that is, it changes its sign if $\nu / \eta$ varies and passes through $\nu / \eta = 1$.

The occurrence of magnetohydrodynamic turbulence does not automatically imply non--zero correlations
of $\bu$ and $\bj$, or $\bomega$ and $\bb$.
It depends on the special circumstances whether, e.g.,  $\lan \bu \cdot \bj \ran$
or $\lan \bomega \cdot \bb \ran$ are different from zero and what their signs are.
For this and other reasons more work is needed to explore the importance
of the $\lan \bu \cdot \bj \ran$ effect in specific settings.
In general the $\lan \bu \cdot \bj \ran$ effect is accompanied by the $\lan \bu \cdot \bb \ran$ effects.
It has then also to be investigated which of these effects dominates.

At first glance there seems to be a inconsistency of our result concerning the $\lan \bu \cdot \bj \ran$ effect
in so far as $\bscE^{(0)}$ should not depend on the choice of the frame of reference but $\bmU$ obviously does.
We must however keep in mind that we have fixed the frame of reference in our calculation by assuming
that there isotropic turbulence occurs in the limit $\bmU \to \bzo$.
When estimating the $\lan \bu \cdot \bj \ran$ effect we have therefore to specify $\bmU$ as the mean velocity of the fluid
relative to the frame in which the assumed causes of turbulence (in simulations the forcing) would,
in the absence of mean motion, just lead to isotropic turbulence.

There is still another issue which has to be considered when applying our result
concerning the $\lan \bu \cdot \bj \ran$ effect to a specific situation.
The deviation of the turbulence from isotropy due to the homogeneous part of the mean motion
is crucial for that effect.
It has to be scrutinized whether such a deviation indeed occurs under the considered circumstances.
The sometimes assumed ``Galilean invariance" of the turbulence (e.g., Sridhar \& Subramanian 2009),
that is, its independence of that part of the mean motion, would exclude the $\lan \bu \cdot \bj \ran$ effect.

In principle we could have determined $\bscE^{(0)}$ in a frame in which $\bmU$ vanishes.
An anisotropy of the turbulence, as calculated in the frame used above,
would lead to a non--vanishing contribution $\bscE^{(00)}$ instead of $c_U \bmU$,
which then has to be considered as another description of the $\lan \bu \cdot \bj \ran$ effect.

\acknowledgements
This work was supported in part by
the European Research Council under the AstroDyn Research Project 227952
and the Swedish Research Council grant 621-2007-4064.

\appendix

\section{Derivation of relation (\ref{eq47}) for $\hat{Q}_{jk}$}
\label{robsow}

Start from $Q_{jk} (\bx, t; \bxi, \tau)$ as given in (\ref{eq45}) and introduce there
the Fourier representations of $\hu_j$ and $\hb_k$ so that
\begin{eqnarray}
Q_{jk} (\bx, t; \bxi, \tau) &=&
\nonumber\\
&& \!\!\!\!\!\!\!\!\!\!\!\!\!\!\!\!\!\!\!\!\!\!\!\!\!\!\!\!\!\!\!\!\!\!\!\!\!\!\!\!\!\!\!\!\!\!\!
    \int \!\!\! \int \int \!\!\! \int \lan \hu_j (\bk^\dag, \omega^\dag) \, \hb_k (\bk^\ddag, \omega^\ddag) \ran
\nonumber\\
&& \!\!\!\!\!\!\!\!\!\!\!\!\!\!\!\!\!\!\!\!\!\!\!
    \exp \big( \iu \big((\bk^\dag + \bk^\ddag) \cdot \bx + (\bk^\dag - \bk^\ddag) \cdot \bxi / 2
\label{eqA01}\\
&& \!\!\!\!\!\!
    - (\omega^\dag + \omega^\ddag) t  -  (\omega^\dag - \omega^\ddag) \tau / 2 \big) \big)
\nonumber\\
&& \qquad \qquad \qquad \quad
    \dd^3 k^\dag \, \dd \omega^\dag  \, \dd^3 k^\ddag \, \dd \omega^\ddag \, .
\nonumber
\end{eqnarray}
Change then the integration variables according to
\begin{eqnarray}
&& \bk^\dag = \bk + \bk'/2 \, , \quad \bk^\ddag = - \bk + \bk'/2
\nonumber\\
&& \omega^\dag = \omega + \omega'/2 \, , \quad \omega^\ddag = - \omega + \omega'/2
\label{eqA03}
\end{eqnarray}
and find so
\begin{eqnarray}
Q_{jk} (\bx, t; \bxi, \tau) &=&
\nonumber\\
&& \!\!\!\!\!\!\!\!\!\!\!\!\!\!\!\!\!\!\!\!\!\!\!\!\!\!\!\!\!\!\!\!\!\!\!\!\!\!\!\!\!\!\!\!\!\!\!
    \int \!\!\! \int \int \!\!\! \int \lan \hu_j (\bk + \bk'/2, \omega + \omega'/2)
\nonumber\\
&& \!\!\!\!\!\!\!\!\!\!\!\!\!\!\!\!\!\!\!\!
    \hb_k (- \bk + \bk'/2, - \omega + \omega'/2) \ran
\nonumber\\
&& \!\!\!\!\!\!\!\!\!\!\!\!\!\!
    \exp \big( \iu (\bk' \cdot \bx - \omega' t) \big) \, \dd^3 k' \, \dd \omega'
\label{eqA05}\\
&& \!\!\!
   \exp \big( \iu (\bk \cdot \bxi - \omega \tau) \big) \, \dd^3 k \, \dd \omega \, .
\nonumber
\end{eqnarray}
This shows that $\hat{Q}_{jk}$ given by (\ref{eq47}) is indeed the Fourier transform
of $Q_{jk} (\bx, t; \bxi, \tau)$ with respect to  $\bxi$ and $\tau$.

\end{document}